%% file: main.tex
\pdfoutput=1
\documentclass[sigconf,nonacm]{acmart}

\usepackage{booktabs}     
\usepackage{enumitem}
\usepackage{multirow}    
\usepackage{tabularx}    
\usepackage{subcaption}  
\usepackage{colortbl}    
\usepackage{siunitx}     

\title{MMQ: Multimodal Mixture-of-Quantization Tokenization for Semantic ID Generation and User Behavioral Adaptation}

\author{Yi Xu}
\affiliation{
  \institution{Alibaba Group}
  \city{Beijing}\country{China}
}
\email{xy397404@alibaba-inc.com}

\author{Moyu Zhang}
\affiliation{
  \institution{Alibaba Group}
  \city{Beijing}\country{China}
}
\email{zhangmoyu.zmy@alibaba-inc.com}

\author{Chenxuan Li}
\affiliation{
  \institution{Peking University}
  \city{Beijing}\country{China}
}
\email{kongocat@stu.pku.edu.cn}

\author{Zhihao Liao}
\affiliation{
  \institution{Beijing University of Aeronautics and Astronautics}
  \city{Beijing}\country{China}
}
\email{liaozhihao@buaa.edu.cn}

\author{Haibo Xing}
\affiliation{
  \institution{Alibaba Group}
  \city{Beijing}\country{China}
}
\email{xinghaibo.xhb@alibaba-inc.com}

\author{Hao Deng}
\affiliation{
  \institution{Alibaba Group}
  \city{Beijing}\country{China}
}
\email{denghao.deng@alibaba-inc.com}

\author{Jinxin Hu}
\affiliation{
  \institution{Alibaba Group}
  \city{Beijing}\country{China}
}
\email{jinxin.hjx@alibaba-inc.com}
\authornote{Corresponding author}

\author{Yu Zhang}
\affiliation{
  \institution{Alibaba Group}
  \city{Beijing}\country{China}
}
\email{daoji@lazada.com}

\author{Xiaoyi Zeng}
\affiliation{
  \institution{Alibaba Group}
  \city{Beijing}\country{China}
}
\email{yuanhan@taobao.com}

\author{Jing Zhang}
\affiliation{
  \institution{Wuhan University, School of Computer Science}
  \city{Wuhan}\country{China}
}
\email{jingzhang.cv@gmail.com}

\begin{document}

\input{secs/00_abstract}


\maketitle

\input{secs/01yl_intro}
\input{secs/02_related_work}

\input{secs/03_method_v2}

\input{secs/04_exp_finalv2}

\input{secs/05_conclusion}

\bibliographystyle{ACM-Reference-Format}
\bibliography{main}

\end{document}

%% file: secs/00_abstract.tex
\begin{abstract}

Recommender systems traditionally represent items using unique identifiers (ItemIDs), but this approach struggles with large, dynamic item corpora and sparse long-tail data, limiting scalability and generalization. Semantic IDs, derived from multimodal content such as text and images, offer a promising alternative by mapping items into a shared semantic space, enabling knowledge transfer and improving recommendations for new or rare items. However, existing methods face two key challenges: (1) balancing cross-modal synergy with modality-specific uniqueness, and (2) bridging the semantic-behavioral gap, where semantic representations may misalign with actual user preferences.
To address these challenges, we propose Multimodal Mixture-of-Quantization (MMQ), a two-stage framework that trains a novel multimodal tokenizer. First, a shared-specific tokenizer leverages a multi-expert architecture with modality-specific and modality-shared experts, using orthogonal regularization to capture comprehensive multimodal information. Second, behavior-aware fine-tuning dynamically adapts semantic IDs to downstream recommendation objectives while preserving modality information through a multimodal reconstruction loss.
Extensive offline experiments and online A/B tests demonstrate that MMQ effectively unifies multimodal synergy, specificity, and behavioral adaptation, providing a scalable and versatile solution for both generative retrieval and discriminative ranking tasks.


\end{abstract}

%% file: secs/01yl_intro.tex
\section{Introduction}
In recommender systems, a common approach represents items using their unique identifiers (ItemIDs) \cite{singh2024bettergeneralizationsemanticids, zheng2025enhancingembeddingrepresentationstability}. However, this paradigm faces critical limitations in real-world scenarios with large, dynamic item corpora. First, high item turnover and shifting popularity trends challenge the stability and scalability of static ID-based embeddings \cite{Milojevi__2010, 10.1145/2523813}. Second, data sparsity for long-tail items leads to poorly generalized representations \cite{Li_2021, ni2023contentdrivenmicrovideorecommendationdataset}, collectively limiting model performance and scalability. These issues motivate exploring alternative item representations.

Semantic IDs, generated from item content such as text and images via multimodal tokenizers, provide a promising alternative \cite{rajput2023recommendersystemsgenerativeretrieval, zheng2025enhancingembeddingrepresentationstability, singh2024bettergeneralizationsemanticids}. By mapping items to a shared semantic space, they enable knowledge transfer among similar items, improving recommendations for new or long-tail items. Moreover, representing items through a combination of tokens from a fixed-size semantic vocabulary enhances scalability and efficiency compared to one-to-one ID mappings \cite{kudo2018subwordregularizationimprovingneural}. These advantages have facilitated semantic IDs' application across diverse recommendation frameworks, from generative retrival \cite{wang2024eagertwostreamgenerativerecommender, hou2023learningvectorquantizeditemrepresentation, Deng_2025} to discriminative ranking tasks \cite{Wang_2025}.

\begin{figure}[t]
\includegraphics[width=0.5\textwidth]{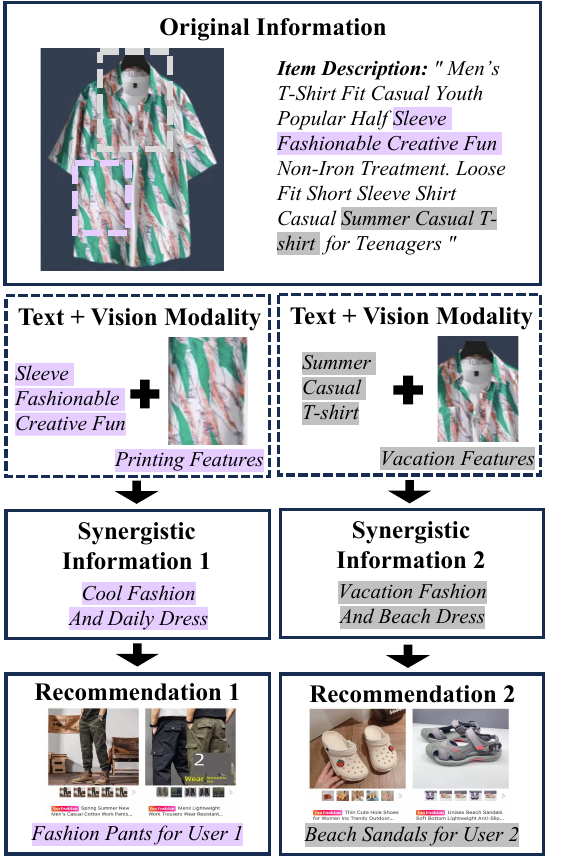}
\centering
\caption{
Illustration of text and vision modal interaction.
}
\vspace{-0.9cm}
\label{fig:realexp}
\end{figure}

Existing semantic tokenizers primarily leverage text and vision modalities, which provide complementary information. According to Partial Information Decomposition (PID) theory \cite{wollstadt2023rigorousinformationtheoreticdefinitionredundancy, liang2023quantifyingmodelingmultimodal}, item information can be decomposed into modality-unique and synergistic components. Capturing both is crucial: synergistic cross-modal interactions can reveal fine-grained item characteristics that single modalities cannot, while modality-unique details preserve unique signals essential for personalization. As illustrated in Figure \ref{fig:realexp}, cross-modal interactions between text and vision reveal finer-grained item characteristics: the left appeals to fashion-forward users, whereas the right attracts casual, vacation-oriented users. Properly modeling this synergy alongside modality-unique details enables more precise recommendations \cite{liu2018efficientlowrankmultimodalfusion,xin2025i2moeinterpretablemultimodalinteractionaware, yu2024mmoeenhancingmultimodalmodels}, such as suggesting fashion pants for the first user and beach sandals for the second.

However, current methods struggle with a trade-off between capturing synergy and uniqueness. Approaches fall into two paradigms: (1) Modality Alignment (MA) merges modalities into a unified representation before tokenization, which may obscure modal-specific details due to dominance of one modality; (2) Modality Separation (MS) maintains separate tokenizers per modality, preserving uniqueness but missing cross-modal synergy. Consequently, generating high-quality semantic IDs that fully leverage multimodal information remains a challenge.
Another key challenge is the \textit{semantic-behavioral gap}: semantic IDs are trained in the semantic space, whereas recommendation relies on user behavior. Purely semantic representations can misrepresent behavioral preferences \cite{wang2024eagertwostreamgenerativerecommender}. For example, semantically similar items may elicit distinct user interactions, leading to noisy recommendations. Thus, semantic tokenizers must adapt dynamically to downstream tasks rather than remain static.

To address these challenges, we propose Multimodal Mixture-of-Quantization (MMQ), which is a two-stage framework. 1) Multimodal Shared-Specific Tokenizer Training: We introduce a multi-expert architecture with modality-specific and modality-shared experts to explicitly model both unique and synergistic information. In addition, we leverage orthogonal regularization to encourage expert diversity, preventing collapse and ensuring comprehensive representation. 2) Behavior-Aware Fine-Tuning: We adapt semantic ID clusters dynamically using downstream recommendation objectives, bridging the semantic-behavioral gap. A multimodal reconstruction loss is proposed to preserve modality information while fine-tuning. We conduct extensive offline experiments and online A/B tests to evaluate MMQ in both generative retrieval and discriminative ranking tasks.

Our main contributions are summarized as follows:
\begin{itemize}
    \item 
    We propose the first unified semantic ID framework that simultaneously captures multimodal synergy and uniqueness while dynamically adapting to user behavior.
    
    \item We design a multi-expert architecture with orthogonal regularization to disentangle modality-shared and modality-specific information. We introduce a behavior-aware fine-tuning to align semantic IDs with downstream recommendation objectives, which effectively addressed the semantic-behavioral gap.
    
    \item Extensive experiments across generative and discriminative tasks demonstrate MMQ's effectiveness, scalability, and versatility.
\end{itemize}

%% file: secs/02_related_work.tex
\section{Related Work}

Recent efforts in item representation have centered on semantic indexing \cite{jin2024languagemodelssemanticindexers,luo2024qarmquantitativealignmentmultimodal, kuai2024breakinghourglassphenomenonresidual, zheng2025pretraininggenerativerecommendermultiidentifier,yang2025sparsemeetsdenseunified}, which generates discrete item identifiers (i.e., tokens) from content information. A dominant approach within this paradigm is to apply quantization or clustering techniques to pre-trained textual embeddings. In particular, TIGER \cite{rajput2023recommendersystemsgenerativeretrieval} is a pioneering work that introduces Residual Quantization (RQ-VAE) to quantize items into semantic IDs, enabling new items to share knowledge through semantic similarity without the pre-trained embeddings. SPM-SID \cite{singh2024bettergeneralizationsemanticids} and PMA \cite{zheng2025enhancingembeddingrepresentationstability} explored the semantic IDs and subwords of Large Language Models (LLMs), using n-gram and SPM(SentencePiece Model) to combine the IDs. Existing works use different quantization methods for tokenization. RQ-Kmeans better addresses the collapse of RQVAE and is widely applied in many works \cite{deng2025onerecunifyingretrieverank}. Unlike the ordered tokenization of RQVAE, unordered tokenization \cite{hou2025generatinglongsemanticids} supports faster parallel generation. VQ-Rec \cite{hou2023learningvectorquantizeditemrepresentation} uses OPQ (Optimized Product Quantization) \cite{jegou2010product} to split and disorderly tokenize text vectors from BERT. ActionPiece \cite{hou2025actionpiececontextuallytokenizingaction} adopts unordered feature sets and context-aware token merging.

Recognizing the limitations of the single content modality, recent research \cite{liu2025generativerecommenderendtoendlearnable} explored the integration of collaborative information into the tokenization. A representative work is EAGER \cite{wang2024eagertwostreamgenerativerecommender}, which combines semantic information with collaborative signals to generate more informative tokens. LC-Rec \cite{zheng2024adaptinglargelanguagemodels} designs multitask semantic alignment fine-tuning, including sequence item prediction, explicit index-language alignment, and implicit alignment tasks. ColaRec \cite{Wang_2024} utilizes a pre-trained collaborative filtering model to incorporate collaborative signals from user-item interactions into hierarchical K-means clustering \cite{huang2025augmentnotcomparativestudy}. IDGenRec \cite{tan2024idgenrecllmrecsysalignmenttextual} converts user behavior sequences into text descriptions and uses the LLMs to generate unique semantic IDs for target items.



OneRec \cite{deng2025onerecunifyingretrieverank} and UTGRec \cite{zheng2025universalitemtokenizationtransferable} are modality-aligned paradigm that perform multimodal embedding alignment on the images and texts obtained from the video. Then, the unified embedding is input to RQ-Kmeans for tokenization. EGA \cite{zheng2025egav2endtoendgenerativeframework} is a modality-separated paradigm that quantizes POI and image embeddings using RQ-VAE, respectively. Then it feeds them into the encoder-decoder for interaction. Consequently, the existing multimodal semantic tokens have the following problems: one must either capture synergy at the risk of losing uniqueness \cite{bai2025chimecompressiveframeworkholistic,liu2025llmalignmentlivestreamingrecommendation}, or preserve uniqueness at the cost of ignoring synergy \cite{yuan2023recommendersystemsidvs}. Existing works that combine these strategies with RQ-VAE or PQ inevitably meet these limitations. To our knowledge, no prior work has proposed a unified tokenizer architecture that can explicitly and simultaneously model both the synergy and uniqueness of multimodal information while mitigating the foundational issues of quantization.

%% file: secs/03_method_v2.tex
\section{Methodology}

\begin{figure*}[htbp]
  \includegraphics[width=1\textwidth]{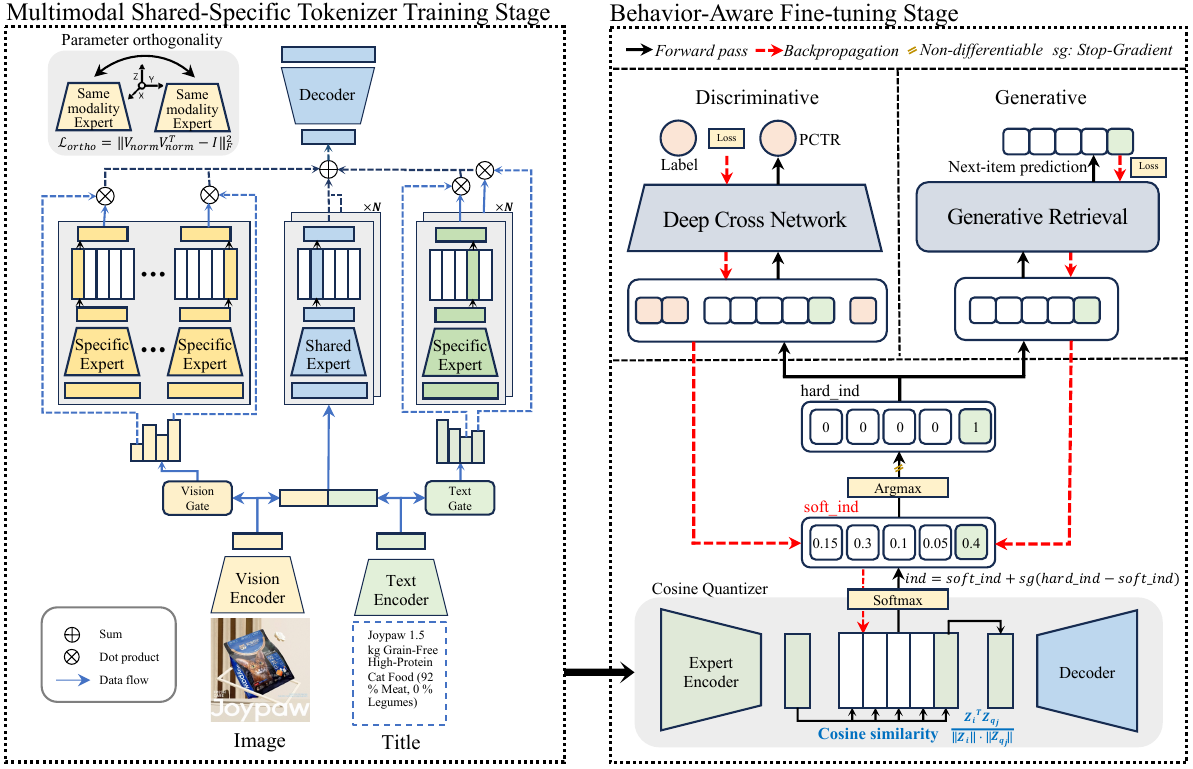}
  \caption{Our Proposed MMQ Framework. (1) Multimodal Shared-Specific Tokenizer Training Stage: We introduce a multi-expert architecture with modality-specific and modality-shared experts to explicitly model both unique and synergistic information. In addition, we leverage orthogonal regularization to encourage expert diversity, preventing expert collapse and ensuring comprehensive representation. (2) Behavior-Aware Fine-Tuning Stage: We adapt semantic ID clusters dynamically using downstream recommendation objectives, bridging the semantic-behavioral gap. }
  \label{fig:fig_intro}
\end{figure*}
In this section, we introduce the proposed item tokenizer framework, termed MMQ, which generates a tuple of semantic IDs for multimodal representations of items, such as text and vision representations. Firstly, we formulate the task of item tokenizer in Section 3.1. Secondly, we introduce the multimodal shared-specific tokenizer, which uses shared and specific experts to explicitly disentangle and model both modality-unique and synergistic information. Thirdly, we introduce the behavior-aware fine-tuning, which guides the dynamic adjustment of cluster centroids using downstream recommendation objectives to achieve alignment between item semantics and user behavioral patterns.
\subsection{Problem Formulation}
The item tokenizer is designed to transform the multimodal content of each item into a sequence of discrete semantic IDs. Formally, for a given item, we first leverage powerful pretrained vision and text embedding models to encode its visual and textual information into continuous feature vectors, respectively, $\mathbf{e}=[\mathbf{e}_{t},\mathbf{e}_{v}]$, where $\mathbf{e}_{t}$ is the pretrained text embedding, $\mathbf{e}_{v}$ is the pretrained vision embedding.  The item tokenizer $\mathcal{T}_{\text{item}}$ maps these high-dimensional embeddings to a discrete sequence of semantic IDs: \begin{equation}
\label{eq:item_tokenization} 
  Semantic\_IDs = (c_1, c_2, \dots, c_l) = \mathcal{T}_{\text{item}}([\mathbf{e}_{t},\mathbf{e}_{v}]) 
\end{equation}
where $l$ is the length of the semantic IDs, $\{c_i\}_i^l$ is the i-th semantic ID.
To ensure that these semantic IDs faithfully preserve the synergistic and unique information of the original text and vision modality, the modality-shared and modality-specifc experts are designed to explicitly disentangle and model both modality-unique and synergistic information.
\subsection{Multimodal Shared-Specific Tokenizer Training Stage}

To balance the synergy and uniqueness of multimodal information, we designed a multimodal shared-specific tokenizer. Furthermore, we added an orthogonal regularization component to address redundancy and replaced the standard L2 distance with cosine similarity during codebook lookup to mitigate the impact of differences in the output value distribution of different modality sub-models.

\subsubsection{modality-shared and modality-specificscalability Expert}
We first introduce the multimodal shared-specific tokenizer, which consists of two core components: modality-specific experts and modality-shared experts. The former is designed to preserve unique, modality-specific properties, while the latter aims to capture synergistic information arising from multimodal interactions.

Specifically, for the $i$-th modality-shared expert $E_{s,i}$ in $\{E_{s,i}\}_{i=1}^{N_s}$, the latent modality-shared representation is $z_{s,i}$.  
\begin{equation}
    \mathbf{z}_{s,i}=E_{s,i}([\mathbf{e}_{t},\mathbf{e}_{v}])
\end{equation}
where $i\in[1,N_s]$ is the $i$-th modality-shared encoders, $N_s$ is the number of modality-shared experts. 

Conversely, the modality-specific experts operate on unimodal inputs, where each expert is dedicated to encoding the unique attributes inherent to a single modality. For the textual modality, a set of $N_t$ dedicated experts, $\{E_{t,i}\}_{i=1}^{N_t}$, transforms the input vector $\mathbf{e}_t$  and $\mathbf{e}_v$ into a corresponding set of latent representations $\{\mathbf{z}_{t,i}\}_{i=1}^{N_t}$. Analogously, the visual modality is processed by the set of $N_v$ vision experts to produce the representations $\{\mathbf{z}_{v,i}\}_{i=1}^{N_v}$. 
\begin{equation}
    \mathbf{z}_{t,i}=E_{t,i}(\mathbf{e}_{t}),\mathbf{z}_{v,i}=E_{v,i}(\mathbf{e}_{v})
\end{equation}

In addition to learning synergistic information across modalities, modality-shared experts also serve a purpose: reducing redundancy between experts. Traditional strategies can cause different experts to learn overlapping information, leading to inefficient parameter usage and consequently redundant expert parameters. However, if shared experts are dedicated to capturing and integrating common knowledge across different modalities, parameter redundancy between other specific experts is reduced. This reduced redundancy contributes to building a more parameter-efficient model with more specialized experts.

For modality-shared experts, the inputs from each modality are deterministically assigned to them. modality-specificscalability experts are dynamically weighted via a gating network that learns to assign an importance score to each expert based on the inputs. 
This fusion mechanism is formally defined as:
\begin{equation}
\label{eq:fusion_mechanism}
\mathbf{z} = \sum_{i=1}^{N_s} \mathbf{z}_{s,i} + \sum_{i=1}^{N_v} g_{v,i} \mathbf{z}_{v,i} + \sum_{i=1}^{N_t} g_{t,i} \mathbf{z}_{t,i}
\end{equation}
\begin{equation}
g_{t}=softmax(MLP_t(\mathbf{e}_{t})+b_t)
\end{equation}
\begin{equation}
g_{v}=softmax(MLP_v(\mathbf{e}_{v})+b_v)\\    
\end{equation}
where $\mathbf{z}_{s,i}$, $\mathbf{z}_{v,i}$, and $\mathbf{z}_{t,i}$ are the latent representations from the modality-shared, visual-specific, and text-specific experts, respectively. The terms $g_{v,i}$ and $g_{t,i}$ represent the learned gating weights.

\textbf{Cosine Distance for Semantic-Aware Quantizer}
To reduce scale mismatches across modality-specific encoders, we replace the standard $L_2$ distance with cosine similarity in codebook lookup. This makes assignments depend on angle rather than magnitude, better matching the semantic geometry of multimodal embeddings and yielding more meaningful token selections. Concretely, for the set of modality-shared latent vectors $\{z_{s,i}\}_{i=1}^{N_s}$, the code index is:
\begin{equation}
c_{s,i}
=\underset{j\in\{1,\dots,K\}}{\arg\max}\;
\frac{\mathbf z_{s,i}^\top \mathbf z_{q,j}}
{\lVert \mathbf z_{s,i}\rVert\,\lVert \mathbf z_{q,j}\rVert},
\qquad i=1,\dots,N_s,
\end{equation}
The cosine distance quantizer captures how well each codeword aligns with $ \mathbf{z}_i $ in direction, independent of magnitude. The searched codeword is denoted as $\{\mathbf{z}_{q_{s,i}}\}_{i=1}^{N_s}$. 

Similarly, for the modality-specific latent vectors $\{\mathbf{z}_{t,i}\}_{i=1}^{N_t}$, $\{\mathbf{z}_{v,i}\}_{i=1}^{N_v}$, the corresponding codebook indexs are selected as $\{c_{t,i}\}_{i=1}^{N_t}$,$\{c_{v,i}\}_{i=1}^{N_v}$. Analogously, the fusion of the quantized representations $z_q$ is formally defined as follows.
\begin{equation}
\mathbf{z}_q = \sum_{j=1}^{N_s} \mathbf{z}_{q_{s,j}} + \sum_{j=1}^{N_v} g_{v,j}·\mathbf{z}_{q_{v,j}} + \sum_{j=1}^{N_t} g_{t,j}·\mathbf{z}_{q_{t,j}}
\end{equation}

\subsubsection{Orthogonal-Constrained Vector-Qantized Code Learning}
\label{section322}To ensure that the generated semantic IDs faithfully preserve both the synergistic and unique information from the original multimodal inputs, we design a dual training objective. First, we perform high-fidelity reconstruction of the original multimodal representations, encouraging the model to capture comprehensive, cross-modal semantics shared across modalities.
Second, we add an orthogonality regularizer on the modality-specific experts, which promotes disentanglement and reduces redundancy across experts, thereby preserving modality-unique cues.

\textbf{Multimodal Reconstruction Loss }
The $\mathcal{L}_{recon}$ is designed to measure the fidelity of the original multimodal embeddings $\mathbf{e}=[\mathbf{e}_t,\mathbf{e}_v]$.
\begin{equation}
\label{eq_recon}
    \mathcal{L}_{recon}=||\mathbf{e}-decoder(\mathbf{z}+sg(\mathbf{z}_q-\mathbf{z}))||^2
\end{equation}
To facilitate the learning process of the modality-specific latent representations, the auxiliary model-specific reconstruction loss is designed in Eq~\eqref{eq_aux}. 
\begin{multline}
\label{eq_aux}
    \mathcal{L}_{aux}=||\mathbf{e}_t-decoder_t(\sum_j^{N_t}(\mathbf{z}_{t,j} +sg(\mathbf{z}_{\mathbf{q}_{t,j}}-\mathbf{z}_{q_{t,j}})))||^2\\
    +||\mathbf{e}_v-decoder_v(\sum_j^{N_t}(\mathbf{z}_{v,j}+sg(\mathbf{z}_{q_{v,j}}-\mathbf{z}_{v,j}))||^2
\end{multline}

\textbf{Orthogonal Regularization for Multi Experts}
In addition to alleviating redundancy implicitly by using shared experts, we also explicitly introduce orthogonality constraints to solve this problem. 
Formally, for a set of $s$ modality-shared experts with learnable weight $\{\mathbf{W}_i\}_{i=1}^N$, 
 \( \mathbf{W}_i \in \mathbb{R}^{d \times d} \) denote the weight matrix of the \( i \)-th expert (\( i = 1, \dots, N \)). We flatten each \( \mathbf{W}_i \) into a vector \( \mathbf{v}_i \in \mathbb{R}^{d^2} \), forming a matrix \( \mathbf{V} = [\mathbf{v}_1, \dots, \mathbf{v}_N] \in \mathbb{R}^{N \times d^2} \). To ensure orthogonality, we normalize each row of \( \mathbf{V} \) to unit length, yielding \( \mathbf{V}_{\text{norm}} \). The orthogonal regularization loss of the group of modality-shared experts is then defined as:
\begin{equation}
\mathcal{L}_{\text{ortho\_shared}} = \left\| \mathbf{V}_{\text{norm}} \mathbf{V}_{\text{norm}}^\top - \mathbf{I} \right\|_F^2,
\end{equation}
where \( \mathbf{I} \) is the identity matrix and \( \|\cdot\|_F^2 \) denotes the squared Frobenius norm. 
Similarly, the orthogonal regularization loss of the group of modality-specificscalability encoders is then defined as:
\begin{equation}
\mathcal{L}_{\text{ortho\_specific}} = \left\| \mathbf{V'}_{\text{norm}} \mathbf{V'}_{\text{norm}}^\top - \mathbf{I} \right\|_F^2,
\end{equation}
where $\mathbf{V'}$ is the flattened weight matrix of modality-specific encoders. The total orthogonal regularization loss is as follows.
\begin{equation}
\mathcal{L}_{\text{ortho}}=\mathcal{L}_{\text{ortho\_shared}}+\sum_i^m
\mathcal{L}_{\text{ortho\_specific}}
\end{equation}
This regularization encourages the expert projections to span distinct directions in the latent space and ensures that multiple experts capture complementary aspects of the input, which is critical for generating compact and semantically rich semantic IDs.

To sum up, the overall training objective is formulated as follows.
\begin{equation}
    \mathcal{L}=\alpha* \mathcal{L}_{recon}+\beta*\mathcal{L}_{aux}+\gamma*\mathcal{L}_{ortho}
\end{equation} 
where $\alpha$, $\beta$ and $\gamma$ are loss weights, unless otherwise stated, we set $\alpha$=12, $\beta$=10, and $\gamma$=0.005 in our experiments. The $\mathcal{L}_{recon}$ is the reconstruction loss to rebuild the original multimodal embeddings $\mathbf{e}=[\mathbf{e}_t,\mathbf{e}_v]$, formally defined in Eq~\eqref{eq_recon}. 


\subsection{Behavior-aware Fine-tuning Stage}
Prevailing semantic ID-based recommenders, such as VQ-REC \cite{hou2023learningvectorquantizeditemrepresentation}  and TIGER \cite{rajput2023recommendersystemsgenerativeretrieval}, operate on a two-stage training pipeline. They first pre-train an item tokenizer to map items into discrete semantic IDs, and then freeze this tokenizer while training downstream recommendation models using the generated IDs. However, this disjointed process creates a critical bottleneck: the discrete, non-differentiable nature of the codebook indices severs the gradient flow, leading to a fundamental misalignment between the tokenizer's semantic space and the downstream model's behavioral objective. To address this, we propose a novel behavior-aware fine-tuning framework that replaces the discrete index lookup with a differentiable, "soft" indexing mechanism. This mechanism formulates item representations as a weighted combination of codebook vectors, thereby creating a continuous path for gradient propagation. This enables the joint optimization of the tokenizer and the recommendation model, effectively aligning the semantic space with the behavioral task while simultaneously preserving the rich, pre-trained knowledge within the codebook.

\textbf{Soft Indices Bridges Item Tokenizer and Downstream tasks} Inspired by Index Backpropagation Quantization (IBQ) \cite{shi2025scalableimagetokenizationindex}, we adapt a soft quantization strategy that computes the cosine similarity logits between each latent embedding and the full codebook. For an item $x$ with visual and textual features, the generated semantic IDs are deconted as $\{\mathbf{c}_i\}_i^l$, the encoded latent representation is $\{\mathbf{z}_i\}_i^l$. 
For each latent represebtation $ \mathbf{z}_i $, we compute its cosine similarity with all codewords $ \mathbf{C}=\{\mathbf{z}_{q_j}\}_{j=1}^K $ in the codebook to form a similarity vector:
\begin{equation}
\mathbf{p}_i = \left[ \frac{\mathbf{z}_i^\top \mathbf{z}_{q_1}}{\|\mathbf{z}_i\| \cdot \|\mathbf{z}_{q_1}\|},\, \dots,\, \frac{\mathbf{z}_i^\top \mathbf{z}_{q_K}}{\|\mathbf{z}_i\| \cdot \|\mathbf{z}_{q_K}\|} \right] \in \mathbb{R}^K.
\end{equation}
The soft indices is obtained by softmax as follows:
\begin{align}
& soft\_ind= \text{softmax}(\mathbf{p}_i/\tau) \in \mathbb{R}^K \\
& hard\_ind = \arg\max_j \frac{\mathbf{z}_i^\top \mathbf{z}_{q_j}}{\|\mathbf{z}_i\| \cdot \|\mathbf{z}_{q_j}\|}  \in \mathbb{R}^K \\
& ind=soft\_ind+sg(hard\_ind-soft\_ind) 
\end{align}
where $ soft\_ind $ represents the probability distribution over the codebook for $ \mathbf{z}_i $, $\tau $ is the temperature coefficient, $sg$ represents stopping gradient. 
The temperature coefficient $ \tau $ controls the sharpness of the soft indices during backpropagation,  where a larger $ \tau $ results in smoother gradients and more stable training, and a smaller $ \tau $ encourages sharper, more discrete behavior. 

\textbf{Joint Optimization Training Object}.
With the Straight-Through Estimator (STE) strategy, during forward propagation, the downstream recommendation tasks use the discrete $hard\_ind$ to look up embeddings for semantic IDs for item representation. During backpropagation, gradients flow through $soft\_ind$, which are differentiable with respect to the item tokenizer's parameters.

This mechanism allows the item tokenizer to be jointly optimized with the downstream recommendation tasks. To preserve the semantic knowledge of the item tokenizer, the joint optimization training object is as follows:
\begin{align}
&\mathcal{L}_{finetune}=\mathcal{L}_{downstream_{task}}+\alpha'* \mathcal{L}_{recon}+\beta'*\mathcal{L}_{aux} 
\end{align}
where $\hat{v}_{t+1}$ is the predicted next item vector, $\alpha'$ and $\beta'$ are hyperparameters to balance the downstream retrieval task and the reconstruction object of the item tokenizer, $\alpha'=0.5$,  $\beta'=0.5$ is set in experiments.

%% file: secs/04_exp_finalv2.tex
\section{Experiments}
In this paper, we conduct extensive experiments on both industrial and public datasets to evaluate the effectiveness of our proposed framework and address the following questions: 
\begin{itemize}[leftmargin=*]
\item \textbf{RQ1}: How does MMQ compare to state-of-the-art item tokenizers in terms of reconstruction accuracy and downstream performance in generative and discriminative recommendation?
\item \textbf{RQ2}: What is the contribution of each component in MMQ, including the modality shared-specific tokenizer and behavior-aware fine-tuning framework to overall performance? 
\item \textbf{RQ3}: How effective is our proposed MMQ in improving recommendations for items with varying degrees of popularity, especially for those in the long-tail items?
\item \textbf{RQ4}: To what extent can the principle of behavior-aware fine-tuning be generalized to other vector quantization methods to enhance their recommendation performance?
\item \textbf{RQ5}: How does the performance of our proposed MMQ scale with an increasing number of semantic IDs?
\item \textbf{RQ6}: To what extent does the modality-shared expert contribute to building a more parameter-efficient model?
\end{itemize}
\subsection{Experimental Setup}
\subsubsection{Dataset} We evaluate the proposed framework both on an industrial dataset and a public dataset. 

\textbf{Industrial Dataset}: This dataset was collected from a leading e-commerce advertising platform in Southeast Asia between October 2024 and May 2025, encompassing 30 million users and 40 million advertisements. It contains user behavior sequences (impressions, clicks, conversions) with an average length of 128, alongside rich, multimodal item content (images, titles, descriptions, etc.). Its scale and complexity make it an ideal benchmark for evaluating real-world performance.

\textbf{Public Dataset}: We conduct experiments on the subset “Beauty” of the Amazon Product Reviews \cite{He_2016} to evaluate our approach, including . All these datasets comprise user review data from May 1996 to September 2014. For the evaluation in generative retrieval, we apply the 5-core filter to exclude unpopular users and items with fewer than five interaction records. Then, we construct user behavior sequences according to the chronological order and uniformly set the maximum item sequence length to 20. For the evaluation in discriminative ranking, we use the user-item rating dataset from Amazon Product Reviews. Following the standard practice, we treat ratings greater than 3 as positive labels and those less than or equal to 3 as negative labels. The dataset is sorted chronologically, and the first 90\% of the data are used for training, while the remaining 10\% constitute the test set.

\subsubsection{Evaluation Metrics}
We employ the quantization metrics to evaluate the performance of the item tokenizer and the recommendation metrics to evaluate the performance of the generated semantic IDs incorporated with the generative retrieval and discriminative ranking task.

\textbf{Quantization Metrics}: Reconstruction Loss \cite{deng2025onerecunifyingretrieverank} is utilized to evalute the reconstrction fidelity for the origin input vector. Codebook utilization \cite{zhu2024scalingcodebooksizevqgan} is employed to reflect the efficiency with which the model uses the codebook vectors. Token Distribution Entropy \cite{bentz2016wordentropynaturallanguages} is utilized to evaluate the diversity and balance of the distribution across semantic codewords in codebooks.

\begin{table*}[htbp]
\centering
\caption{Overall performance comparison on two datasets. We evaluate all methods on two downstream tasks: discriminative ranking and generative retrieval. Best results in each column are in \textbf{bold}. Our model, MMQ, is highlighted in gray. The last row (Improv.) denotes the relative improvement of MMQ over the best baseline.}
\label{tab:main_comparison}

\begin{subtable}{\textwidth}
    \centering 
    \caption{Discriminative Ranking Evaluation}
    \label{tab:discriminative_results}
    \footnotesize 
    \begin{tabular}{l | ccc|cc | ccc|cc}
        \toprule
        \multirow{2}{*}{\textbf{Methods}} & \multicolumn{5}{c|}{\textbf{Industrial Dataset}} & \multicolumn{5}{c}{\textbf{Amazon Beauty}} \\
        \cmidrule(lr){2-6} \cmidrule(lr){7-11}
        & $L_{\text{recon}}\downarrow$ & Entropy$\uparrow$ & Util.$\uparrow$ & AUC$\uparrow$ & GAUC$\uparrow$ & $L_{\text{recon}}\downarrow$ & Entropy$\uparrow$ & Util.$\uparrow$ & AUC$\uparrow$ & GAUC$\uparrow$ \\
        \midrule
        PPNet & - & - & - & 0.7144 & 0.6034 & - & - & - & 0.6455 & 0.5897 \\
        \midrule
        MA-RQ-VAE & 0.6165 & 7.1178 & 0.79 & 0.7169 & 0.6064 & \underline{0.6028} & 3.4904 & 0.99 & 0.6466 & 0.5952 \\
        MA-RQ-kmeans & 0.6759 & \underline{7.7933} & \underline{1.00} & 0.7170 & 0.6056 & 0.6240 & 1.7100 & \underline{1.00} & 0.6472 & 0.5999 \\
        MA-OPQ & 0.6730 & 5.3092 & 0.33 & 0.7149 & 0.6046 & 0.9647 & 3.3980 & 0.96 & 0.6469 & 0.5998 \\
        \midrule
        MS-RQ-VAE & \underline{0.5547} & 6.3389 & 0.99 & 0.7171 & 0.6067 & 1.1001 & 3.8375 & \underline{1.00} & 0.6473 & 0.5971 \\
        MS-RQ-kmeans & 0.6848 & 7.5962 & \underline{1.00} & \underline{0.7181} & \underline{0.6076} & 1.1547 & \underline{4.1603} & \underline{1.00} & \underline{0.6480} & \underline{0.6001} \\
        MS-OPQ & 0.5678 & 5.2789 & 0.30 & 0.7177 & 0.6071 & 0.9801 & 3.3308 & 0.95 & \underline{0.6481} & 0.5995 \\
        \midrule
        \rowcolor{gray!15}
        \textbf{MMQ (Ours)} & \textbf{0.5529} & \textbf{7.9560} & \textbf{1.00} & \textbf{0.7184} & \textbf{0.6081} & \textbf{0.4470} & \textbf{4.4206} & \textbf{1.00} & \textbf{0.6503} & \textbf{0.6020} \\
        \textit{Improv.} & \textit{+0.32\%} & \textit{+2.08\%} & \textit{+0.00\%} & \textit{+0.04\%} & \textit{+0.07\%} & \textit{+25.84\%} & \textit{+6.25\%} & \textit{+0.00\%} & \textit{+0.33\%} & \textit{+0.31\%} \\
        \bottomrule
    \end{tabular}
\end{subtable}

\vspace{1em} %

\begin{subtable}{\textwidth}
    \centering
    \caption{Generative Retrieval Evaluation}
    \label{tab:generative_results}
    \resizebox{\textwidth}{!}{%
    \begin{tabular}{l | ccc|cccc | ccc|cccc}
        \toprule
        \multirow{2}{*}{\textbf{Methods}} & \multicolumn{7}{c|}{\textbf{Industrial Dataset}} & \multicolumn{7}{c}{\textbf{Amazon Beauty}} \\
        \cmidrule(lr){2-8} \cmidrule(lr){9-15}
        & $L_{\text{recon}}\downarrow$ & Entropy$\uparrow$ & Util.$\uparrow$ & R@5$\uparrow$ & R@10$\uparrow$ & N@5$\uparrow$ & N@10$\uparrow$ & $L_{\text{recon}}\downarrow$ & Entropy$\uparrow$ & Util.$\uparrow$ & R@5$\uparrow$ & R@10$\uparrow$ & N@5$\uparrow$ & N@10$\uparrow$ \\
        \midrule
        MA-RQ-VAE & 0.6165 & 7.1178 & 0.79 & 0.0570 & 0.0754 & 0.1362 & 0.1621 & \underline{0.6028} & 3.4904 & 0.99 & 0.0343 & 0.0477 & 0.0236 & 0.0292 \\
        MA-RQ-kmeans & 0.6759 & \underline{7.7933} & \underline{1.00} & 0.0488 & 0.0522 & 0.1339 & 0.1388 & 0.6240 & 1.7100 & \underline{1.00} & 0.0344 & 0.0467 & 0.0242 & 0.0294 \\
        MA-OPQ & 0.6730 & 5.3092 & 0.33 & 0.0223 & 0.0264 & 0.0561 & 0.0619 & 0.9647 & 3.3980 & 0.96 & 0.0324 & 0.0407 & 0.0218 & 0.0250 \\
        \midrule
        MS-RQ-VAE & \underline{0.5547} & 6.3389 & 0.99 & \underline{0.0779} & \underline{0.0988} & \underline{0.1897} & \underline{0.2191} & 1.1001 & 3.8375 & \underline{1.00} & 0.0274 & 0.0399 & 0.0188 & 0.0241 \\
        MS-RQ-kmeans & 0.6848 & 7.5962 & \underline{1.00} & 0.0497 & 0.0534 & 0.1359 & 0.1411 & 1.1547 & \underline{4.1603} & \underline{1.00} & \underline{0.0413} & \underline{0.0549} & \underline{0.0293} & \underline{0.0349} \\
        MS-OPQ & 0.5678 & 5.2789 & 0.30 & 0.0757 & 0.0927 & 0.1866 & 0.2105 & 0.9801 & 3.3308 & 0.95 & 0.0411 & 0.0517 & 0.0280 & 0.0322 \\
        \midrule
        \rowcolor{gray!15}
        \textbf{MMQ (Ours)} & \textbf{0.5529} & \textbf{7.9560} & \textbf{1.00} & \textbf{0.1034} & \textbf{0.1192} & \textbf{0.2661} & \textbf{0.2883} & \textbf{0.4470} & \textbf{4.4206} & \textbf{1.00} & \textbf{0.0455} & \textbf{0.0675} & \textbf{0.0296} & \textbf{0.0384} \\
        \textit{Improv.} & \textit{+0.32\%} & \textit{+2.08\%} & \textit{+0.00\%} & \textit{+32.73\%} & \textit{+20.64\%} & \textit{+40.27\%} & \textit{+31.58\%} & \textit{+25.84\%} & \textit{+6.25\%} & \textit{+0.00\%} & \textit{+10.16\%} & \textit{+22.95\%} & \textit{+1.02\%} & \textit{+10.02\%} \\
        \bottomrule
    \end{tabular}
    } 
\end{subtable}

\end{table*}

\textbf{Recommendation Metrics}: In generative retrieval, Recall@N and NDCG@N with N=5,10 are used to evaluate the performace. In discriminative ranking, AUC and GAUC are used to evaluate the performance. For the online experiments, the Adervertising Revenue, CVR (Conversion Rate), and Order are used to evaluate the online performace.

\subsubsection{Baselines} To provide a comprehensive comparison and align with the existing literature, we structure our experiments around two dominant paradigms for multimodal semantic ID generation: \textbf{Modality-Aligned (MA) and Modality-Separated (MS).} Under the two paradigms, we experiment with the various quantization methods employed in semantic ID generation for recommendation tasks. We select the currently mainstream quantization methods based on semantic ID generation as our baseline.

\textbf{RQ-VAE}: Residual-Quantized Variational AutoEncoder (RQ-VAE) \cite{zeghidour2021soundstreamendtoendneuralaudio} is a multi-level vector quantization model. It generates a coarse-to-fine sequence of semantic IDs by sequentially quantizing the residual of the input representation at each level. Due to its ability to produce hierarchical semantic structures, RQ-VAE is widely adopted for large-scale generative retrieval and discriminative ranking tasks.

\textbf{RQ-Kmeans}: Similar to RQ-VAE, RQ-Kmeans \cite{luo2024qarmquantitativealignmentmultimodal} also employs a residual quantization scheme to generate semantic IDs in a coarse-to-fine manner. Instead of learning the codebooks via backpropagation, RQKmeans constructs them heuristically by applying K-means clustering directly to the representation residuals at each stage. This combination of residual quantization and K-means can lead to more stable training dynamics for semantic ID generation.

\textbf{OPQ}: Optimized Product Quantization (OPQ) is an advanced variant of Product Quantization (PQ) \cite{jegou2010product}. It partitions the input representation into a set of disjoint sub-vectors and quantizes each one independently using a dedicated codebook. A key characteristic of OPQ is that it produces an unordered set of semantic IDs (one ID per sub-vector), which offers greater flexibility for downstream recommendation models\cite{hou2025generatinglongsemanticids,hou2023learningvectorquantizeditemrepresentation}. 
\subsubsection{Experiment Setup}

\textbf{Recommendation Foundations}: For the evaluation of the Generative Retrieval task, we adopt HeterRec \cite{Deng_2025}, a strong multi-token prediction model, as our base framework. For the Discriminative Ranking task, we employ the well-established Parameter Personalized Network (PPNet) \cite{chang2023pepnetparameterembeddingpersonalized} as the backbone architecture.

\textbf{Implementation Details}: For the item tokenizer training, in the industrial dataset, the codebook size is set to 3,000. IN public datasets, the codebook size is set to 100. The length of the semantic ID sequence is consistently set to 6 for all settings, specifically, $N_s=2$, $N_t=2$ and $N_v=2$ are set for MMQ. For both the industrial dataset and public dataset, the text embedding is obtained from the pre-trained Qwen3-Embedding 7B \cite{zhang2025qwen3embeddingadvancingtext} with the dimension of 256, and the vision embedding is obtained from the pre-trained Pailitao v8 with the dimension of 256.

\begin{table*}
\centering
\captionsetup{width=0.9\textwidth, justification=centering} 
\caption{Ablation Experiments.}
\label{tab:performance}
\begin{tabular}{l|ccccccc}
\toprule
\multirow{2}{*}{Methods} & \multicolumn{7}{c}{Generative Retrieval Evaluation} \\ 
\cmidrule(lr){2-8}
&$\mathbf{L}_{recon}$ & Entropy & Utilization
 & R@5 & R@10&N@5 & N@10
\\ 
\midrule
MMQ &0.5529&7.9560 &1.00 &0.1034&0.1192& 0.2661&0.2883\\
w/o Cosine Quantizer &0.5817 &4.9524  &0.59  & 0.0786&0.0965&0.1936&0.2188 \\
w/o Auxiliary Reconstruction Loss & 0.5879&4.3854 &0.51 &0.0684&0.0834&0.1695&0.1907\\
w/o Orthogonal Regularization  & 0.5816 & 7.5640 &0.68  &0.0583&0.0717&0.1111&0.1330 \\
w/o Behavior-Aware Fine-tuning &0.5726  &7.6927 &0.99&0.0792&0.0997& 0.1606&0.1931\\
\bottomrule

\end{tabular}
\end{table*}

\subsection{Overall Performance (RQ1)}
As presented in Table~\ref{tab:main_comparison}, we conduct a comprehensive evaluation on both Generative Retrieval and Discriminative Ranking tasks. We compare our proposed method, MMQ, against various quantization methods under two distinct multimodal fusion paradigms: Modality-Aligned and Modality-Separated. The analysis focuses on both the reconstruction fidelity and the downstream recommendation performance.

\textbf{Comparison with Modality-Separated(MS) and Modality-Aligned(MA) Fusion}:
The Modality-Separated approach consistently outperforms the Modality-Aligned approach on the large-scale industrial dataset across both ranking and retrieval tasks. On the smaller Amazon Beauty dataset, the performance gap between the two is less pronounced. This suggests that as the item vocabulary and data complexity grow, forcing different modalities into a single, shared embedding space significantly increases quantization difficulty. The encoder struggles to distill effective, generalized information from the aligned representation, even if the codebook utilization appears high. Consequently, the downstream performance of modality-aligned methods can be constrained on large-scale data, highlighting the importance of preserving modality-specific characteristics, which is a core principle of our MMQ design.

\textbf{Effectiveness of the Proposed MMQ}: MMQ consistently surpasses both Modality-Aligned and Modality-Separated baseline strategies across all tasks and datasets. Specifically, MMQ achieves the lowest reconstruction loss and maintains high codebook utilization. More importantly, the semantic IDs generated by MMQ yield superior accuracy in both GR and Discriminative Ranking tasks. 
The reduced reconstruction error indicates that our multimodal shared-specific expert architecture effectively captures both shared and distinct features, leading to more robust and stable representation learning. This structural advantage directly translates to higher-quality semantic IDs. The state-of-the-art downstream performance decisively validates the efficacy of our proposed multimodal shared-specific Tokenizer and, crucially, the Behavior-aware Fine-tuning stage. This demonstrates that by explicitly modeling synergistic and unique information and then aligning it with downstream objectives, we can generate semantic IDs that are not only representative but also highly effective for recommendations.

\textbf{Comparison Semantic IDs with Traditional Item IDs}: In the discriminative ranking task, we observe a consistent trend across both industrial and public datasets: methods leveraging learned semantic IDs generally outperform the traditional item ID-based backbone (PPNet). This foundational finding underscores the inherent advantage of using rich, multifaceted semantic representations over opaque, monolithic identifiers. It strongly motivates the core premise of our work, as learning a high-quality item tokenizer is a crucial step toward building more advanced and effective recommendation systems.
\subsection{Ablation Study (RQ2)}
We conduct the ablation experiments in Table \ref{tab:performance} to study how each module contributes to the overall performance of MMQ.

\textbf{The Impact of Orthogonal Regularization}
As shown in Table \ref{tab:performance}, ablating the orthogonality constraint leads to a sharp decline in both codebook utilization and token distribution entropy. This phenomenon indicates that without the explicit pressure to learn distinct features, the multi-expert network is susceptible to expert collapse, where different experts fail to specialize and instead learn redundant, overlapping information. This lack of specialization introduces significant redundancy among the generated semantic IDs, which in turn degrades the performance on downstream recommendation tasks. This finding underscores the criticality of the orthogonality constraint for ensuring representational diversity and overall model effectiveness.

\begin{figure}[htbp]
  \centering
  \begin{subfigure}[t]{\linewidth}
    \centering
    \includegraphics[width=0.95\linewidth]{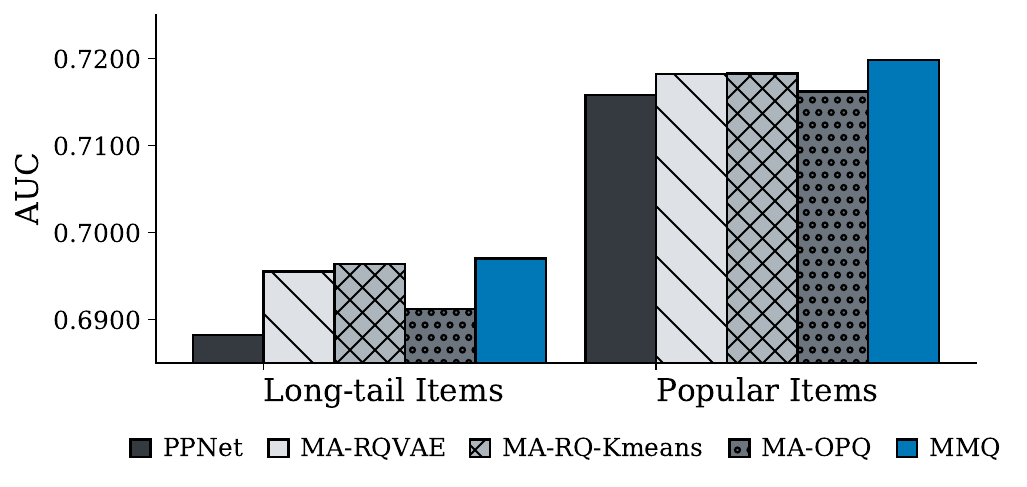}
    \caption{Comparison on Modality-Separated (MS) paradigms.}
    \label{fig:top}
  \end{subfigure}
  \vspace{1em} %
  \begin{subfigure}[t]{\linewidth}
    \centering
    \includegraphics[width=0.95\linewidth]{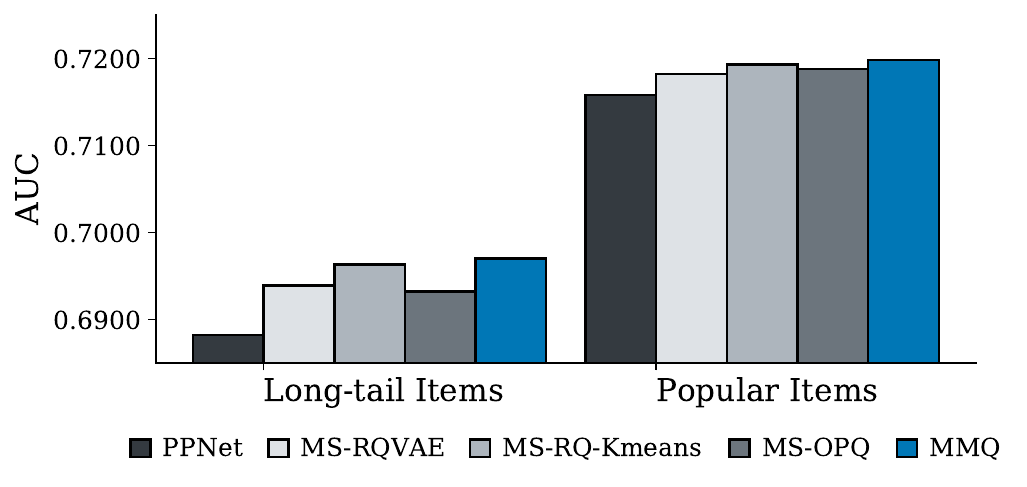}
    \caption{Comparison on Modality-Aligned (MA) paradigms.}
    \label{fig:bottom}
  \end{subfigure}
 \caption{Item Popularity Stratified Performance Comparison.}
  \label{fig:Longtail}
  \vspace{-0.3cm}
\end{figure}
\textbf{The Impact of Auxiliary Model-Specific Reconstruction Loss}: When the modality-specific auxiliary loss is removed, both codebook utilization and reconstruction loss degrade significantly. This indicates that the auxiliary model-specific reconstruction loss plays a crucial role in guiding the model-specific encoders to learn more discriminative representations and promoting the modality-shared encoder to learn common characteristics.

\textbf{The Impact of Cosine Quantizer}: Comparing the L2 distance and cosine distance in multimodal vector quantization, the cosine distance improves both codebook utilization and token distribution entropy, which demonstrates that angular proximity-based quantizer brings a more uniform and diverse use of the codebook.

\textbf{The Impact of Behavior-aware Fine-tuning}: Removing the behavior-aware fine-tuning strategy results in a consistent drop in both Recall and NDCG across all settings. This indicates an inherent inconsistency between the semantic space and the behavior domain, which hinders the effective transfer of semantic information from the item space to the user behavior space, resulting in degraded downstream recommendation performance. Behavior-aware fine-tuning addresses this issue by aligning the item-semantic ID mapping with the downstream task objective, thereby enabling more effective semantic representation adaptation for behavioral prediction.

\subsection{Item Popularity Stratified Performance Analysis (RQ3)}
A key hypothesis of our work is that semantic IDs enhance generalization by enabling knowledge sharing among semantically similar items, a benefit particularly crucial for long-tail items suffering from data sparsity. To validate this, we stratify items from the industrial dataset based on their popularity (accumulated impressions) into Popular Items (Top 25\%) and Long-tail Items (Bottom 25\%). We then evaluate different semantic ID generation methods under both the Modality-Aligned and Modality-Separated  paradigms. As depicted in Figure.\ref{fig:Longtail}, a consistent trend emerges: all methods utilizing semantic IDs achieve significant improvements in AUC on long-tail items compared to what a traditional Item ID-based model would yield. This performance lift is observed across both MA and MS strategies. Notably, our proposed MMQ delivers the most substantial gains in this challenging scenario. The pronounced improvement in the long-tail cohort confirms that semantic IDs effectively mitigate the data sparsity issue. The superior performance of MMQ further indicates that it learns a more  meaningful item-to-semantic mapping relationship. This mapping relationship is more adept at capturing the nuanced semantic relationships that are vital for recommending items with limited interaction data, thereby validating the effectiveness of our model design.

\subsection{The Compatibility of the Behavior-aware Fine-tuning Framework (RQ4)}
\begin{figure}[t]
\includegraphics[width=0.3\textwidth]{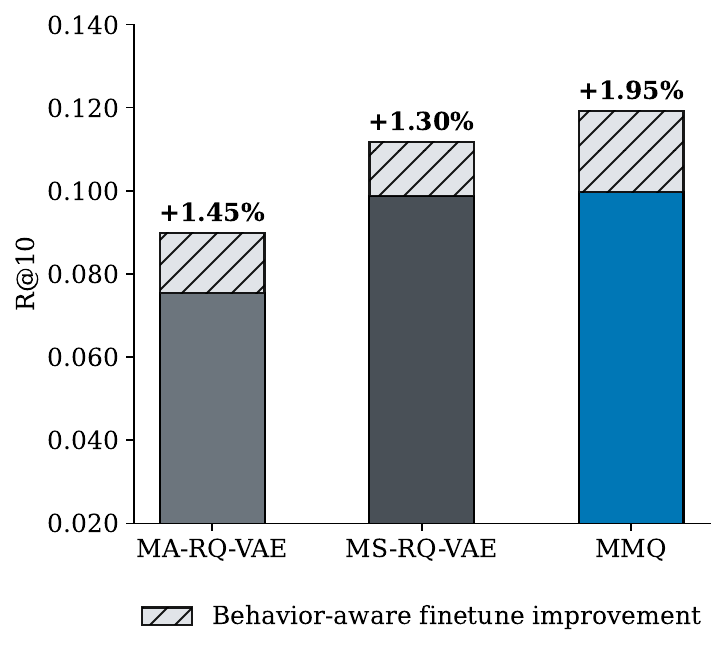}
\centering
\caption{
The Compatibility Experiments on Integrating Behavior-Aware Fine-tuning with RQ-VAE.
}
\vspace{-0.2cm}
\label{fig:finetune_improvement}
\end{figure}

To assess the generalizability and plug-and-play nature of our proposed Behavior-aware Fine-tuning, we integrate it into a widely-adopted baseline, RQ-VAE, under both the Modality-Aligned  and Modality-Separated paradigms. As shown in Figure.\ref{fig:finetune_improvement},  applying BAF yields consistent and significant downstream performance gains for the RQ-VAE model, irrespective of the underlying fusion paradigm. This finding validates two critical points. First, it confirms that behavior-aware fine-tuning is a model-agnostic enhancement, demonstrating strong compatibility and portability across different vector quantization methods. Second, its consistent success highlights its core capability: effectively bridging the well-known content-behavior gap by dynamically adjusting cluster centroids using downstream recommendation objectives to achieve alignment between item semantics and user behavioral patterns.


\begin{figure}[t]
\includegraphics[width=0.45\textwidth]{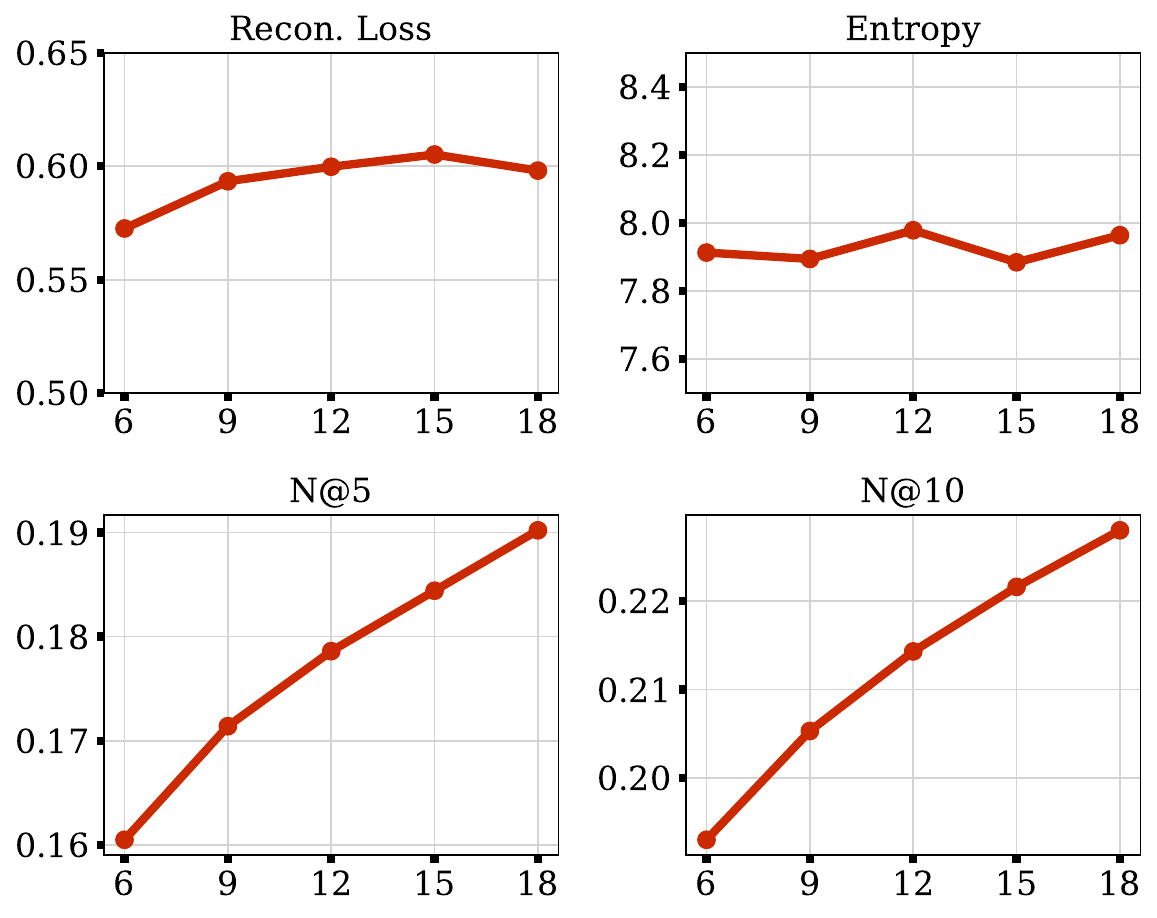}
\centering
\caption{
The scalability of the semantic ID length.
}
\vspace{-0.2cm}
\label{fig:scalability}

\end{figure}

\subsection{The Scalability of Semantic ID Length (RQ5)}
The length of the semantic ID sequence is a critical hyperparameter that determines the granularity of item representation. We investigate its impact by varying the length across the set of [6, 9, 12, 15, 18]. As depicted in Figure.\ref{fig:scalability}, we observe a distinct positive correlation: downstream recommendation accuracy, measured by NDCG@5 and NDCG@10, consistently improves as the sequence length increases. Crucially, this performance gain is achieved without compromising representation quality; both the reconstruction loss and the token distribution entropy remain stable across all settings. This dual-faceted result—enhanced recommendation performance coupled with stable quantization fidelity—demonstrates that our MMQ framework scales effectively with an increasing number of semantic IDs. It can leverage longer semantic sequences to capture progressively richer item semantics without degrading training stability or representation quality.

\subsection{Experiment Analysis on the Contribution of Modality-Shared Experts (RQ6)}
\begin{figure}[t]
\includegraphics[width=0.45\textwidth]{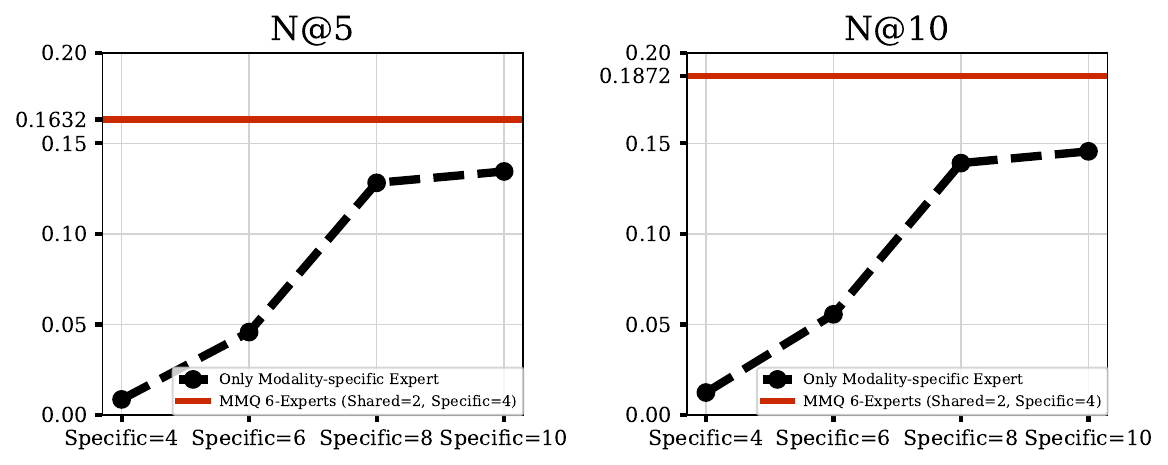}
\centering
\caption{
Experiment Analysis on the Contribution of
Modality-Shared Experts
}
\vspace{-0.2cm}
\label{fig:scalability}
\end{figure}

In this experiment, we investigate whether the shared expert effectively mitigates redundancy and promotes to build a more parameter-efficient model with more specialized experts. We compare our proposed model, MMQ (Shared=2,Specific=4), against four "specific-only" variants that lack the shared component, where the number of specific experts are set to [4,6,8,10].
As shown in Figure \ref{fig:scalability}, the Specific-Only (4 experts) model performs substantially worse than the MMQ (Shared=2, Specific=4). More strikingly, even the Specific-Only (10 experts) variant, despite its significantly larger capacity, still fails to match the performance of the MMQ (Shared=2, Specific=4). This suggests that without a shared expert to handle common knowledge, specific experts are tent to learn redundant, overlapping information, leading to inefficient parameter utilization. This experiment has demonstrated the shared expert is crucial for building a more parameter-efficient model with highly specialized experts.

\section{Online Experiments}
To conclusively validate the real-world efficacy of our proposed method, we deployed the generated semantic IDs into the generative retrieval system of a large-scale e-commerce platform. A rigorous online A/B test was conducted over a 30-day period (July 6–August 6, 2025). The experimental group, which received recommendations based on our semantic IDs, was allocated 10\% of the platform's random user traffic, while the control group continued to use the existing Item ID-based system. The online experiment yielded significant and consistent improvements across key business metrics: \textbf{0.90\% increase in Advertising Revenue,
+4.33\%  increase in Conversion Rate (CVR),
3.52\% increase in Orders}.
These substantial online gains robustly confirm the practical value and effectiveness of our approach, demonstrating its readiness for deployment in production-scale recommendation systems.

%% file: secs/05_conclusion.tex
\section{Conclusion}

In this paper, we presented a multimodal mixture-of-quantization (MMQ) that integrates a multi-expert architecture with both modality-shared and modality-specific experts. Through a two-stage training framework: multimodal shared-specific pretraining followed by behavior-aware fine-tuning, MMQ fuses heterogeneous modalities into compact semantic IDs while explicitly aligning the semantic space with the behavioral space of recommendation tasks. This design effectively resolves the mismatch between content-derived and behavior-driven representations. Extensive evaluations on industrial datasets, public benchmarks, and online A/B tests demonstrate that MMQ consistently outperforms state-of-the-art baselines in both generative retrieval and discriminative ranking, delivering robust performance gains and strong generalization across domains. Beyond improving recommendation accuracy, MMQ offers a scalable and flexible approach to semantic ID generation, potentially benefiting other multimodal applications such as personalized search, advertising, and cross-domain content understanding.

While promising, our approach has certain limitations. The two-stage framework introduces additional training complexity and may require substantial multimodal data to fully exploit modality-specific experts. Moreover, the behavior-aware fine-tuning stage is task-dependent, which may limit direct transferability to unrelated downstream tasks. Future research could explore lightweight or continual-learning variants of MMQ to reduce computational overhead, extend the architecture to incorporate temporal or interactive modalities (e.g., video, audio, user-session dynamics), and investigate self-supervised or few-shot adaptations for data-scarce settings. We believe that MMQ represents a meaningful step toward bridging the gap between semantic representation learning and user behavior modeling, and we expect it to inspire further research in multimodal recommendation and beyond.


%% file: main.bbl
\begin{thebibliography}{10}
\providecommand{\url}[1]{#1}
\csname url@samestyle\endcsname
\providecommand{\newblock}{\relax}
\providecommand{\bibinfo}[2]{#2}
\providecommand{\BIBentrySTDinterwordspacing}{\spaceskip=0pt\relax}
\providecommand{\BIBentryALTinterwordstretchfactor}{4}
\providecommand{\BIBentryALTinterwordspacing}{\spaceskip=\fontdimen2\font plus
\BIBentryALTinterwordstretchfactor\fontdimen3\font minus \fontdimen4\font\relax}
\providecommand{\BIBforeignlanguage}[2]{{%
\expandafter\ifx\csname l@#1\endcsname\relax
\typeout{** WARNING: IEEEtran.bst: No hyphenation pattern has been}%
\typeout{** loaded for the language `#1'. Using the pattern for}%
\typeout{** the default language instead.}%
\else
\language=\csname l@#1\endcsname
\fi
#2}}
\providecommand{\BIBdecl}{\relax}
\BIBdecl

\bibitem{em:86}
R.~Engelmore and A.~Morgan, Eds., \emph{Blackboard Systems}.\hskip 1em plus 0.5em minus 0.4em\relax Reading, Mass.: Addison-Wesley, 1986.

\bibitem{c:83}
W.~J. Clancey, ``{Communication, Simulation, and Intelligent Agents: Implications of Personal Intelligent Machines for Medical Education},'' in \emph{Proceedings of the Eighth International Joint Conference on Artificial Intelligence {(IJCAI-83)}}.\hskip 1em plus 0.5em minus 0.4em\relax Menlo Park, Calif: {IJCAI Organization}, 1983, pp. 556--560.

\bibitem{c:84}
------, ``{Classification Problem Solving},'' in \emph{Proceedings of the Fourth National Conference on Artificial Intelligence}.\hskip 1em plus 0.5em minus 0.4em\relax Menlo Park, Calif.: AAAI Press, 1984, pp. 45--54.

\bibitem{r:80}
\BIBentryALTinterwordspacing
A.~L. Robinson, ``New ways to make microcircuits smaller,'' \emph{Science}, vol. 208, no. 4447, pp. 1019--1022, 1980. [Online]. Available: \url{https://science.sciencemag.org/content/208/4447/1019}
\BIBentrySTDinterwordspacing

\bibitem{r:80x}
------, ``{New Ways to Make Microcircuits Smaller---Duplicate Entry},'' \emph{Science}, vol. 208, pp. 1019--1026, 1980.

\bibitem{hcr:83}
\BIBentryALTinterwordspacing
D.~W. Hasling, W.~J. Clancey, and G.~Rennels, ``Strategic explanations for a diagnostic consultation system,'' \emph{International Journal of Man-Machine Studies}, vol.~20, no.~1, pp. 3--19, 1984. [Online]. Available: \url{https://www.sciencedirect.com/science/article/pii/S0020737384800036}
\BIBentrySTDinterwordspacing

\bibitem{hcrt:83}
D.~W. Hasling, W.~J. Clancey, G.~R. Rennels, and T.~Test, ``{Strategic Explanations in Consultation---Duplicate},'' \emph{The International Journal of Man-Machine Studies}, vol.~20, no.~1, pp. 3--19, 1983.

\bibitem{r:86}
J.~Rice, ``{Poligon: A System for Parallel Problem Solving},'' Dept.\ of Computer Science, Stanford Univ., Technical Report KSL-86-19, 1986.

\bibitem{c:79}
W.~J. Clancey, ``{Transfer of Rule-Based Expertise through a Tutorial Dialogue},'' {Ph.D.} diss., Dept.\ of Computer Science, Stanford Univ., Stanford, Calif., 1979.

\bibitem{c:21}
------, ``{The Engineering of Qualitative Models},'' 2021, forthcoming.

\bibitem{c:22}
A.~Vaswani, N.~Shazeer, N.~Parmar, J.~Uszkoreit, L.~Jones, A.~N. Gomez, L.~Kaiser, and I.~Polosukhin, ``Attention is all you need,'' 2017.

\bibitem{c:23}
{NASA}, ``Pluto: The 'other' red planet,'' \url{https://www.nasa.gov/nh/pluto-the-other-red-planet}, 2015, accessed: 2018-12-06.

\bibitem{zhai2024actionsspeaklouderwords}
\BIBentryALTinterwordspacing
J.~Zhai, L.~Liao, X.~Liu, Y.~Wang, R.~Li, X.~Cao, L.~Gao, Z.~Gong, F.~Gu, M.~He, Y.~Lu, and Y.~Shi, ``Actions speak louder than words: Trillion-parameter sequential transducers for generative recommendations,'' 2024. [Online]. Available: \url{https://arxiv.org/abs/2402.17152}
\BIBentrySTDinterwordspacing

\bibitem{ji2023genreclargelanguagemodel}
\BIBentryALTinterwordspacing
J.~Ji, Z.~Li, S.~Xu, W.~Hua, Y.~Ge, J.~Tan, and Y.~Zhang, ``Genrec: Large language model for generative recommendation,'' 2023. [Online]. Available: \url{https://arxiv.org/abs/2307.00457}
\BIBentrySTDinterwordspacing

\bibitem{liu2025generativerecommenderendtoendlearnable}
\BIBentryALTinterwordspacing
E.~Liu, B.~Zheng, C.~Ling, L.~Hu, H.~Li, and W.~X. Zhao, ``Generative recommender with end-to-end learnable item tokenization,'' 2025. [Online]. Available: \url{https://arxiv.org/abs/2409.05546}
\BIBentrySTDinterwordspacing

\bibitem{han2025mtgrindustrialscalegenerativerecommendation}
\BIBentryALTinterwordspacing
R.~Han, B.~Yin, S.~Chen, H.~Jiang, F.~Jiang, X.~Li, C.~Ma, M.~Huang, X.~Li, C.~Jing, Y.~Han, M.~Zhou, L.~Yu, C.~Liu, and W.~Lin, ``Mtgr: Industrial-scale generative recommendation framework in meituan,'' 2025. [Online]. Available: \url{https://arxiv.org/abs/2505.18654}
\BIBentrySTDinterwordspacing

\bibitem{rajput2023recommendersystemsgenerativeretrieval}
\BIBentryALTinterwordspacing
S.~Rajput, N.~Mehta, A.~Singh, R.~H. Keshavan, T.~Vu, L.~Heldt, L.~Hong, Y.~Tay, V.~Q. Tran, J.~Samost, M.~Kula, E.~H. Chi, and M.~Sathiamoorthy, ``Recommender systems with generative retrieval,'' 2023. [Online]. Available: \url{https://arxiv.org/abs/2305.05065}
\BIBentrySTDinterwordspacing

\bibitem{deng2025onerecunifyingretrieverank}
\BIBentryALTinterwordspacing
J.~Deng, S.~Wang, K.~Cai, L.~Ren, Q.~Hu, W.~Ding, Q.~Luo, and G.~Zhou, ``Onerec: Unifying retrieve and rank with generative recommender and iterative preference alignment,'' 2025. [Online]. Available: \url{https://arxiv.org/abs/2502.18965}
\BIBentrySTDinterwordspacing

\bibitem{yang2025sparsemeetsdenseunified}
\BIBentryALTinterwordspacing
Y.~Yang, Z.~Ji, Z.~Li, Y.~Li, Z.~Mo, Y.~Ding, K.~Chen, Z.~Zhang, J.~Li, S.~Li, and L.~Liu, ``Sparse meets dense: Unified generative recommendations with cascaded sparse-dense representations,'' 2025. [Online]. Available: \url{https://arxiv.org/abs/2503.02453}
\BIBentrySTDinterwordspacing

\bibitem{hou2025generatinglongsemanticids}
\BIBentryALTinterwordspacing
Y.~Hou, J.~Li, A.~Shin, J.~Jeon, A.~Santhanam, W.~Shao, K.~Hassani, N.~Yao, and J.~McAuley, ``Generating long semantic ids in parallel for recommendation,'' 2025. [Online]. Available: \url{https://arxiv.org/abs/2506.05781}
\BIBentrySTDinterwordspacing

\bibitem{tang2025modelalllargelanguage}
\BIBentryALTinterwordspacing
Z.~Tang, Z.~Huan, Z.~Li, X.~Zhang, J.~Hu, C.~Fu, J.~Zhou, L.~Zou, and C.~Li, ``One model for all: Large language models are domain-agnostic recommendation systems,'' 2025. [Online]. Available: \url{https://arxiv.org/abs/2310.14304}
\BIBentrySTDinterwordspacing

\bibitem{wang2024learnableitemtokenizationgenerative}
\BIBentryALTinterwordspacing
W.~Wang, H.~Bao, X.~Lin, J.~Zhang, Y.~Li, F.~Feng, S.-K. Ng, and T.-S. Chua, ``Learnable item tokenization for generative recommendation,'' 2024. [Online]. Available: \url{https://arxiv.org/abs/2405.07314}
\BIBentrySTDinterwordspacing

\bibitem{qu2024tokenreclearningtokenizeid}
\BIBentryALTinterwordspacing
H.~Qu, W.~Fan, Z.~Zhao, and Q.~Li, ``Tokenrec: Learning to tokenize id for llm-based generative recommendation,'' 2024. [Online]. Available: \url{https://arxiv.org/abs/2406.10450}
\BIBentrySTDinterwordspacing

\bibitem{singh2024bettergeneralizationsemanticids}
\BIBentryALTinterwordspacing
A.~Singh, T.~Vu, N.~Mehta, R.~Keshavan, M.~Sathiamoorthy, Y.~Zheng, L.~Hong, L.~Heldt, L.~Wei, D.~Tandon, E.~H. Chi, and X.~Yi, ``Better generalization with semantic ids: A case study in ranking for recommendations,'' 2024. [Online]. Available: \url{https://arxiv.org/abs/2306.08121}
\BIBentrySTDinterwordspacing

\bibitem{zheng2025enhancingembeddingrepresentationstability}
\BIBentryALTinterwordspacing
C.~Zheng, M.~Huang, D.~Pedchenko, K.~Rangadurai, S.~Wang, G.~Nahum, J.~Lei, Y.~Yang, T.~Liu, Z.~Luo, X.~Wei, D.~Ramasamy, J.~Yang, Y.~Han, L.~Yang, H.~Xu, R.~Jin, and S.~Yang, ``Enhancing embedding representation stability in recommendation systems with semantic id,'' 2025. [Online]. Available: \url{https://arxiv.org/abs/2504.02137}
\BIBentrySTDinterwordspacing

\bibitem{hou2023learningvectorquantizeditemrepresentation}
\BIBentryALTinterwordspacing
Y.~Hou, Z.~He, J.~McAuley, and W.~X. Zhao, ``Learning vector-quantized item representation for transferable sequential recommenders,'' 2023. [Online]. Available: \url{https://arxiv.org/abs/2210.12316}
\BIBentrySTDinterwordspacing

\bibitem{hou2025actionpiececontextuallytokenizingaction}
\BIBentryALTinterwordspacing
Y.~Hou, J.~Ni, Z.~He, N.~Sachdeva, W.-C. Kang, E.~H. Chi, J.~McAuley, and D.~Z. Cheng, ``Actionpiece: Contextually tokenizing action sequences for generative recommendation,'' 2025. [Online]. Available: \url{https://arxiv.org/abs/2502.13581}
\BIBentrySTDinterwordspacing

\bibitem{wang2024eagertwostreamgenerativerecommender}
\BIBentryALTinterwordspacing
Y.~Wang, J.~Xun, M.~Hong, J.~Zhu, T.~Jin, W.~Lin, H.~Li, L.~Li, Y.~Xia, Z.~Zhao, and Z.~Dong, ``Eager: Two-stream generative recommender with behavior-semantic collaboration,'' 2024. [Online]. Available: \url{https://arxiv.org/abs/2406.14017}
\BIBentrySTDinterwordspacing

\bibitem{zheng2024adaptinglargelanguagemodels}
\BIBentryALTinterwordspacing
B.~Zheng, Y.~Hou, H.~Lu, Y.~Chen, W.~X. Zhao, M.~Chen, and J.-R. Wen, ``Adapting large language models by integrating collaborative semantics for recommendation,'' 2024. [Online]. Available: \url{https://arxiv.org/abs/2311.09049}
\BIBentrySTDinterwordspacing

\bibitem{Wang_2024}
\BIBentryALTinterwordspacing
Y.~Wang, Z.~Ren, W.~Sun, J.~Yang, Z.~Liang, X.~Chen, R.~Xie, S.~Yan, X.~Zhang, P.~Ren, Z.~Chen, and X.~Xin, ``Content-based collaborative generation for recommender systems,'' in \emph{Proceedings of the 33rd ACM International Conference on Information and Knowledge Management}, ser. CIKM ’24.\hskip 1em plus 0.5em minus 0.4em\relax ACM, Oct. 2024, p. 2420–2430. [Online]. Available: \url{http://dx.doi.org/10.1145/3627673.3679692}
\BIBentrySTDinterwordspacing

\bibitem{huang2025augmentnotcomparativestudy}
\BIBentryALTinterwordspacing
W.-H. Huang, C.-W. Ke, W.-N. Chiu, Y.-X. Su, C.-C. Yang, C.-Y. Cheng, Y.-N. Chen, and P.-J. Cheng, ``Augment or not? a comparative study of pure and augmented large language model recommenders,'' 2025. [Online]. Available: \url{https://arxiv.org/abs/2505.23053}
\BIBentrySTDinterwordspacing

\bibitem{tan2024idgenrecllmrecsysalignmenttextual}
\BIBentryALTinterwordspacing
J.~Tan, S.~Xu, W.~Hua, Y.~Ge, Z.~Li, and Y.~Zhang, ``Idgenrec: Llm-recsys alignment with textual id learning,'' 2024. [Online]. Available: \url{https://arxiv.org/abs/2403.19021}
\BIBentrySTDinterwordspacing

\bibitem{jin2024languagemodelssemanticindexers}
\BIBentryALTinterwordspacing
B.~Jin, H.~Zeng, G.~Wang, X.~Chen, T.~Wei, R.~Li, Z.~Wang, Z.~Li, Y.~Li, H.~Lu, S.~Wang, J.~Han, and X.~Tang, ``Language models as semantic indexers,'' 2024. [Online]. Available: \url{https://arxiv.org/abs/2310.07815}
\BIBentrySTDinterwordspacing

\bibitem{zheng2025universalitemtokenizationtransferable}
\BIBentryALTinterwordspacing
B.~Zheng, H.~Lu, Y.~Chen, W.~X. Zhao, and J.-R. Wen, ``Universal item tokenization for transferable generative recommendation,'' 2025. [Online]. Available: \url{https://arxiv.org/abs/2504.04405}
\BIBentrySTDinterwordspacing

\bibitem{zheng2025egav2endtoendgenerativeframework}
\BIBentryALTinterwordspacing
Z.~Zheng, Z.~Wang, F.~Yang, J.~Fan, T.~Zhang, Y.~Wang, and X.~Wang, ``Ega-v2: An end-to-end generative framework for industrial advertising,'' 2025. [Online]. Available: \url{https://arxiv.org/abs/2505.17549}
\BIBentrySTDinterwordspacing

\bibitem{luo2024qarmquantitativealignmentmultimodal}
\BIBentryALTinterwordspacing
X.~Luo, J.~Cao, T.~Sun, J.~Yu, R.~Huang, W.~Yuan, H.~Lin, Y.~Zheng, S.~Wang, Q.~Hu, C.~Qiu, J.~Zhang, X.~Zhang, Z.~Yan, J.~Zhang, S.~Zhang, M.~Wen, Z.~Liu, K.~Gai, and G.~Zhou, ``Qarm: Quantitative alignment multi-modal recommendation at kuaishou,'' 2024. [Online]. Available: \url{https://arxiv.org/abs/2411.11739}
\BIBentrySTDinterwordspacing

\bibitem{kuai2024breakinghourglassphenomenonresidual}
\BIBentryALTinterwordspacing
Z.~Kuai, Z.~Chen, H.~Wang, M.~Li, D.~Miao, B.~Wang, X.~Chen, L.~Kuang, Y.~Han, J.~Wang, G.~Tang, L.~Liu, S.~Wang, and J.~Zhuo, ``Breaking the hourglass phenomenon of residual quantization: Enhancing the upper bound of generative retrieval,'' 2024. [Online]. Available: \url{https://arxiv.org/abs/2407.21488}
\BIBentrySTDinterwordspacing

\bibitem{zheng2025pretraininggenerativerecommendermultiidentifier}
\BIBentryALTinterwordspacing
B.~Zheng, E.~Liu, Z.~Chen, Z.~Ma, Y.~Wang, W.~X. Zhao, and J.-R. Wen, ``Pre-training generative recommender with multi-identifier item tokenization,'' 2025. [Online]. Available: \url{https://arxiv.org/abs/2504.04400}
\BIBentrySTDinterwordspacing

\bibitem{bai2025chimecompressiveframeworkholistic}
\BIBentryALTinterwordspacing
Y.~Bai, R.~Xiang, K.~Li, Y.~Tang, Y.~Cheng, X.~Liu, P.~Jiang, and K.~Gai, ``Chime: A compressive framework for holistic interest modeling,'' 2025. [Online]. Available: \url{https://arxiv.org/abs/2504.06780}
\BIBentrySTDinterwordspacing

\bibitem{yuan2023recommendersystemsidvs}
\BIBentryALTinterwordspacing
Z.~Yuan, F.~Yuan, Y.~Song, Y.~Li, J.~Fu, F.~Yang, Y.~Pan, and Y.~Ni, ``Where to go next for recommender systems? id- vs. modality-based recommender models revisited,'' 2023. [Online]. Available: \url{https://arxiv.org/abs/2303.13835}
\BIBentrySTDinterwordspacing

\bibitem{liu2025llmalignmentlivestreamingrecommendation}
\BIBentryALTinterwordspacing
Y.~Liu, J.~Cao, S.~Wang, S.~Wen, X.~Chen, X.~Wu, S.~Yang, Z.~Liu, K.~Gai, and G.~Zhou, ``Llm-alignment live-streaming recommendation,'' 2025. [Online]. Available: \url{https://arxiv.org/abs/2504.05217}
\BIBentrySTDinterwordspacing

\bibitem{wollstadt2023rigorousinformationtheoreticdefinitionredundancy}
\BIBentryALTinterwordspacing
P.~Wollstadt, S.~Schmitt, and M.~Wibral, ``A rigorous information-theoretic definition of redundancy and relevancy in feature selection based on (partial) information decomposition,'' 2023. [Online]. Available: \url{https://arxiv.org/abs/2105.04187}
\BIBentrySTDinterwordspacing

\bibitem{liang2023quantifyingmodelingmultimodal}
\BIBentryALTinterwordspacing
P.~P. Liang, Y.~Cheng, X.~Fan, C.~K. Ling, S.~Nie, R.~Chen, Z.~Deng, N.~Allen, R.~Auerbach, F.~Mahmood, R.~Salakhutdinov, and L.-P. Morency, ``Quantifying \& modeling multimodal interactions: An information decomposition framework,'' 2023. [Online]. Available: \url{https://arxiv.org/abs/2302.12247}
\BIBentrySTDinterwordspacing

\bibitem{liu2018efficientlowrankmultimodalfusion}
\BIBentryALTinterwordspacing
Z.~Liu, Y.~Shen, V.~B. Lakshminarasimhan, P.~P. Liang, A.~Zadeh, and L.-P. Morency, ``Efficient low-rank multimodal fusion with modality-specific factors,'' 2018. [Online]. Available: \url{https://arxiv.org/abs/1806.00064}
\BIBentrySTDinterwordspacing

\bibitem{xin2025i2moeinterpretablemultimodalinteractionaware}
\BIBentryALTinterwordspacing
J.~Xin, S.~Yun, J.~Peng, I.~Choi, J.~L. Ballard, T.~Chen, and Q.~Long, ``I2moe: Interpretable multimodal interaction-aware mixture-of-experts,'' 2025. [Online]. Available: \url{https://arxiv.org/abs/2505.19190}
\BIBentrySTDinterwordspacing

\bibitem{xue2023dynamicmultimodalfusion}
\BIBentryALTinterwordspacing
Z.~Xue and R.~Marculescu, ``Dynamic multimodal fusion,'' 2023. [Online]. Available: \url{https://arxiv.org/abs/2204.00102}
\BIBentrySTDinterwordspacing

\bibitem{yu2024mmoeenhancingmultimodalmodels}
\BIBentryALTinterwordspacing
H.~Yu, Z.~Qi, L.~Jang, R.~Salakhutdinov, L.-P. Morency, and P.~P. Liang, ``Mmoe: Enhancing multimodal models with mixtures of multimodal interaction experts,'' 2024. [Online]. Available: \url{https://arxiv.org/abs/2311.09580}
\BIBentrySTDinterwordspacing

\bibitem{mustafa2022multimodalcontrastivelearninglimoe}
\BIBentryALTinterwordspacing
B.~Mustafa, C.~Riquelme, J.~Puigcerver, R.~Jenatton, and N.~Houlsby, ``Multimodal contrastive learning with limoe: the language-image mixture of experts,'' 2022. [Online]. Available: \url{https://arxiv.org/abs/2206.02770}
\BIBentrySTDinterwordspacing

\bibitem{jin2024moeacceleratingmixtureofexpertsmethods}
\BIBentryALTinterwordspacing
P.~Jin, B.~Zhu, L.~Yuan, and S.~Yan, ``Moe++: Accelerating mixture-of-experts methods with zero-computation experts,'' 2024. [Online]. Available: \url{https://arxiv.org/abs/2410.07348}
\BIBentrySTDinterwordspacing

\bibitem{deepseekai2025deepseekv3technicalreport}
\BIBentryALTinterwordspacing
DeepSeek-AI, A.~Liu, and a.~Z.~P. Bei Feng~et a, ..., ``Deepseek-v3 technical report,'' 2025. [Online]. Available: \url{https://arxiv.org/abs/2412.19437}
\BIBentrySTDinterwordspacing

\bibitem{wang2024auxiliarylossfreeloadbalancingstrategy}
\BIBentryALTinterwordspacing
L.~Wang, H.~Gao, C.~Zhao, X.~Sun, and D.~Dai, ``Auxiliary-loss-free load balancing strategy for mixture-of-experts,'' 2024. [Online]. Available: \url{https://arxiv.org/abs/2408.15664}
\BIBentrySTDinterwordspacing

\bibitem{huang2025largescalegenerativeranking}
\BIBentryALTinterwordspacing
Y.~Huang, Y.~Chen, X.~Cao, R.~Yang, M.~Qi, Y.~Zhu, Q.~Han, Y.~Liu, Z.~Liu, X.~Yao, Y.~Jia, L.~Ma, Y.~Zhang, T.~Zhu, L.~Zhang, L.~Chen, W.~Chen, M.~Zhu, R.~Xu, and L.~Zhang, ``Towards large-scale generative ranking,'' 2025. [Online]. Available: \url{https://arxiv.org/abs/2505.04180}
\BIBentrySTDinterwordspacing

\bibitem{Wang_2025}
\BIBentryALTinterwordspacing
C.~Wang, B.~Wu, Z.~Chen, L.~Shen, B.~Wang, and X.~Zeng, ``Scaling transformers for discriminative recommendation via generative pretraining,'' in \emph{Proceedings of the 31st ACM SIGKDD Conference on Knowledge Discovery and Data Mining V.2}, ser. KDD ’25.\hskip 1em plus 0.5em minus 0.4em\relax ACM, Aug. 2025, p. 2893–2903. [Online]. Available: \url{http://dx.doi.org/10.1145/3711896.3737117}
\BIBentrySTDinterwordspacing

\bibitem{Deng_2025}
\BIBentryALTinterwordspacing
H.~Deng, H.~Xing, K.~Matsuyama, Y.~Huang, J.~Hu, H.~Wen, J.~Xu, Z.~Chen, Y.~Zhang, X.~Zeng, and J.~Zhang, ``Heterrec: Heterogeneous information transformer for scalable sequential recommendation,'' in \emph{Proceedings of the 48th International ACM SIGIR Conference on Research and Development in Information Retrieval}, ser. SIGIR ’25.\hskip 1em plus 0.5em minus 0.4em\relax ACM, Jul. 2025, p. 3020–3024. [Online]. Available: \url{http://dx.doi.org/10.1145/3726302.3730206}
\BIBentrySTDinterwordspacing

\bibitem{chang2023pepnetparameterembeddingpersonalized}
\BIBentryALTinterwordspacing
J.~Chang, C.~Zhang, Y.~Hui, D.~Leng, Y.~Niu, Y.~Song, and K.~Gai, ``Pepnet: Parameter and embedding personalized network for infusing with personalized prior information,'' 2023. [Online]. Available: \url{https://arxiv.org/abs/2302.01115}
\BIBentrySTDinterwordspacing

\bibitem{shi2025scalableimagetokenizationindex}
\BIBentryALTinterwordspacing
F.~Shi, Z.~Luo, Y.~Ge, Y.~Yang, Y.~Shan, and L.~Wang, ``Scalable image tokenization with index backpropagation quantization,'' 2025. [Online]. Available: \url{https://arxiv.org/abs/2412.02692}
\BIBentrySTDinterwordspacing

\bibitem{Milojevi__2010}
\BIBentryALTinterwordspacing
S.~Milojević, ``Power law distributions in information science: Making the case for logarithmic binning,'' \emph{Journal of the American Society for Information Science and Technology}, vol.~61, no.~12, p. 2417–2425, Dec. 2010. [Online]. Available: \url{http://dx.doi.org/10.1002/asi.21426}
\BIBentrySTDinterwordspacing

\bibitem{10.1145/2523813}
\BIBentryALTinterwordspacing
J.~a. Gama, I.~\v{Z}liobaitundefined, A.~Bifet, M.~Pechenizkiy, and A.~Bouchachia, ``A survey on concept drift adaptation,'' \emph{ACM Comput. Surv.}, vol.~46, no.~4, Mar. 2014. [Online]. Available: \url{https://doi.org/10.1145/2523813}
\BIBentrySTDinterwordspacing

\bibitem{jegou2010product}
H.~Jegou, M.~Douze, and C.~Schmid, ``Product quantization for nearest neighbor search,'' \emph{IEEE transactions on pattern analysis and machine intelligence}, vol.~33, no.~1, pp. 117--128, 2010.

\bibitem{kudo2018subwordregularizationimprovingneural}
\BIBentryALTinterwordspacing
T.~Kudo, ``Subword regularization: Improving neural network translation models with multiple subword candidates,'' 2018. [Online]. Available: \url{https://arxiv.org/abs/1804.10959}
\BIBentrySTDinterwordspacing

\bibitem{ni2023contentdrivenmicrovideorecommendationdataset}
\BIBentryALTinterwordspacing
Y.~Ni, Y.~Cheng, X.~Liu, J.~Fu, Y.~Li, X.~He, Y.~Zhang, and F.~Yuan, ``A content-driven micro-video recommendation dataset at scale,'' 2023. [Online]. Available: \url{https://arxiv.org/abs/2309.15379}
\BIBentrySTDinterwordspacing

\bibitem{pancha2022pinnerformersequencemodelinguser}
\BIBentryALTinterwordspacing
N.~Pancha, A.~Zhai, J.~Leskovec, and C.~Rosenberg, ``Pinnerformer: Sequence modeling for user representation at pinterest,'' 2022. [Online]. Available: \url{https://arxiv.org/abs/2205.04507}
\BIBentrySTDinterwordspacing

\bibitem{Li_2021}
\BIBentryALTinterwordspacing
S.~Li, W.~Lei, Q.~Wu, X.~He, P.~Jiang, and T.-S. Chua, ``Seamlessly unifying attributes and items: Conversational recommendation for cold-start users,'' \emph{ACM Transactions on Information Systems}, vol.~39, no.~4, p. 1–29, Aug. 2021. [Online]. Available: \url{http://dx.doi.org/10.1145/3446427}
\BIBentrySTDinterwordspacing

\bibitem{Ying_2018}
\BIBentryALTinterwordspacing
R.~Ying, R.~He, K.~Chen, P.~Eksombatchai, W.~L. Hamilton, and J.~Leskovec, ``Graph convolutional neural networks for web-scale recommender systems,'' in \emph{Proceedings of the 24th ACM SIGKDD International Conference on Knowledge Discovery amp; Data Mining}, ser. KDD ’18.\hskip 1em plus 0.5em minus 0.4em\relax ACM, Jul. 2018, p. 974–983. [Online]. Available: \url{http://dx.doi.org/10.1145/3219819.3219890}
\BIBentrySTDinterwordspacing

\bibitem{zeghidour2021soundstreamendtoendneuralaudio}
\BIBentryALTinterwordspacing
N.~Zeghidour, A.~Luebs, A.~Omran, J.~Skoglund, and M.~Tagliasacchi, ``Soundstream: An end-to-end neural audio codec,'' 2021. [Online]. Available: \url{https://arxiv.org/abs/2107.03312}
\BIBentrySTDinterwordspacing

\bibitem{He_2016}
\BIBentryALTinterwordspacing
R.~He and J.~McAuley, ``Ups and downs: Modeling the visual evolution of fashion trends with one-class collaborative filtering,'' in \emph{Proceedings of the 25th International Conference on World Wide Web}, ser. WWW ’16.\hskip 1em plus 0.5em minus 0.4em\relax International World Wide Web Conferences Steering Committee, Apr. 2016, p. 507–517. [Online]. Available: \url{http://dx.doi.org/10.1145/2872427.2883037}
\BIBentrySTDinterwordspacing

\bibitem{zhu2024scalingcodebooksizevqgan}
\BIBentryALTinterwordspacing
L.~Zhu, F.~Wei, Y.~Lu, and D.~Chen, ``Scaling the codebook size of vqgan to 100,000 with a utilization rate of 99\%,'' 2024. [Online]. Available: \url{https://arxiv.org/abs/2406.11837}
\BIBentrySTDinterwordspacing

\bibitem{bentz2016wordentropynaturallanguages}
\BIBentryALTinterwordspacing
C.~Bentz and D.~Alikaniotis, ``The word entropy of natural languages,'' 2016. [Online]. Available: \url{https://arxiv.org/abs/1606.06996}
\BIBentrySTDinterwordspacing

\bibitem{zhang2025qwen3embeddingadvancingtext}
\BIBentryALTinterwordspacing
Y.~Zhang, M.~Li, D.~Long, X.~Zhang, H.~Lin, B.~Yang, P.~Xie, A.~Yang, D.~Liu, J.~Lin, F.~Huang, and J.~Zhou, ``Qwen3 embedding: Advancing text embedding and reranking through foundation models,'' 2025. [Online]. Available: \url{https://arxiv.org/abs/2506.05176}
\BIBentrySTDinterwordspacing

\end{thebibliography}
